\documentclass[12pt]{amsart}
\usepackage{amssymb}
\usepackage{amsmath,amsthm,amsfonts}
\usepackage[applemac]{inputenc}  \usepackage{graphicx}
\usepackage{pdfsync}
\usepackage{fancyhdr}
\usepackage[applemac]{inputenc}

\usepackage{color}
\usepackage{xcolor}
\usepackage[breaklinks,colorlinks,backref]{hyperref}

\hypersetup{
    colorlinks, 
    linktoc=all, 
    linkcolor=red,  
  citecolor=hyptxt,
  urlcolor=blue}
  \hypersetup{
  citebordercolor=,
  filebordercolor=green,
  linkbordercolor=blue
}
\definecolor{hyptxt}{rgb}{0.7, 0.4, 0.9}

\usepackage{bibtopic}

\usepackage[full]{textcomp} 
     \usepackage{fbb} 
     \usepackage[scaled=.95]{cabin}
     \usepackage[varqu,varl]{inconsolata}
     \usepackage[libertine,bigdelims,vvarbb]{newtxmath}
     \usepackage[cal=boondoxo]{mathalfa}
    \usepackage[T1]{fontenc}
\useosf

\newtheorem{prop}{Proposition}[section]

\newcommand{\beprop}{\begin{prop}}
\newcommand{\enprop}{\end{prop}}
\newcommand{\bprf}{\begin{proof}}
\newcommand{\eprf}{\end{proof}}

\newcommand{\ket}[1]{|\kern.3ex#1\kern.3ex\rangle}
\newcommand{\bra}[1]{\langle\kern.3ex #1 \kern.3ex|}
\newcommand{\scalar}[2]{\langle\kern.3ex #1 \kern.3ex|\kern.3ex#2\kern.3ex\rangle}

\def\R{\mathbb{R}}
\def\N{\mathbb{N}}
\def\C{\mathbb{C}}
\def\Z{\mathbb{Z}}
\def\D{\mathcal{D}}

\def\lg{\langle }
\def\rg{\rangle }
\def\deq{\stackrel{\mathrm{def}}{=}}

\def\ii{\mathrm{i}}
\def\ud{\mathrm{d}}

\def\sfP{\mathsf{P}}

\def\sfM{\mathsf{M}}

\def\sfAw{\mathsf{A}^{w;\eta}}

\def\MW{\sfM^{w;\eta}}

\def\MWF{\sfM^{w_1;\eta}}
\def\MWS{\sfM^{w_2;\eta}}

\definecolor{hervecolor}{rgb}{0.8,0,0.7}

\numberwithin{equation}{section}

\begin{document}
\date{\today}

\title{Covariant integral  quantization of the unit disk}
\author[Del Olmo, Gazeau]{M.~A. del Olmo$^{\mathrm{a}}$ and
J.P. Gazeau$^{\mathrm{b,c}}$}
\address{\emph{$^{\mathrm{a}}$
Departamento de F\'{\i}sica Te\'orica and IMUVA, Universidad de Valladolid, E-47011, Valladolid, Spain}}
\address{\emph{$^{\mathrm{b}}$ Centro Brasileiro de Pesquisas F\'{\i}sicas } \\
\emph{Rua Xavier Sigaud 150, 22290-180 - Rio de Janeiro, RJ, Brazil  }}
\address{\emph{  $^{\mathrm{c}}$ APC, UMR 7164,}\\
\emph{ Univ Paris  Diderot, Sorbonne Paris Cit\'e,}
\emph{75205 Paris, France}}

\email{e-mail: olmo@fta.uva.es,
gazeau@apc.in2p3.fr}

{\abstract{We implement  a SU$(1,1)$ covariant integral quantization of functions or distributions on the unit disk. The latter can be viewed as the phase space for the motion of a test ``massive'' particle on 1+1 Anti de Sitter space-time, and the relevant unitary irreducible representations of SU$(1,1)$ corresponding to the quantum version of such motions are found in the discrete series and its lower limits. Our quantization method depends on a weight function on the phase space, and it includes Perelomov coherent states (CS) quantization.  
Semi-classical portraits or lower symbols of main physically relevant operators are determined.}}

\maketitle

\tableofcontents

\section{Introduction}
\label{intro}
The group SU$(1,1)$, which is the two-fold covering of SO$_0(1,2)$ can be interpreted as a dynamical group for the $1+1$ Anti-de-Sitter as a space-time  and the unit disk $\mathcal{D}$ as a phase space, i.e. the set of free motions   with a fixed ``energy'' at rest \cite{gazhus92,gazren93A,gazren93B,debievre94}. Therefore, a comprehensive program of quantization of the unit disk as a phase space by using all resources of covariant integral quantization as it is defined for instance in \cite{bergaz14,aagbook13,gazeauAP16,gazmur16} (and references therein) is appealing. This program includes as well the semi-classical return to the original phase space through the construction of the so-called lower (Lieb, \cite{lieb73}) or covariant (Berezin, \cite{berezin74}) symbols, which have a true probabilistic interpretation when the integral quantization is based on normalised positive operator valued measures, simply denoted here by POVM, the latter acronym being used for normalised POVM as well.

The organisation of the paper is as follows. Section \ref{Dgeom}  and \ref{SU11rep}  are a reminder of a well-known material about SU$(1,1)$, as it can found in a classical treatise in group representation theory like \cite{vilenkin68} or in \cite{aagbook13}. In Section \ref{Dgeom} we describe the geometry of the unit disk,  as a Kaelherian manifold in Subsection \ref{Dkaehler}, as a left coset of SU$(1,1)$ in Subsection \ref{Dcoset},  with group action in the usual way, and, in Subsection \ref{Dphasespace},  as a phase space for the motion of a test particle in $1+1$ Anti de Sitter space-time, with the identification of three basic observables as SU$(1,1)$ generators.  In Subsection \ref{AdScoset} we complete these geometric and algebraic aspects  with the description of the $1+1$ AdS space-time as also a left coset of SU$(1,1)$ viewed as the double covering of its kinematical group SO$_0(2,1)$.
Section \ref{SU11rep} is devoted to some representations of SU$(1,1)$ relevant to our purposes. In Subsection \ref{discreteseries} we give a concise description of the discrete series (in a wide sense) of representations of SU$(1,1)$ as acting on Fock-Bargman Hilbert spaces of holomorphic functions in the unit disk, and, in Subsection \ref{su11}, of their generators as first-order differential operators on these functions. 
In Section \ref{covIQ} we first recall the  framework of covariant integral quantization (see for instance \cite{gazeauAP16} and references therein), when it is associated with a unitary irreducible representation (UIR) of a Lie group in Subsection \ref{CIQUIR}, with a square-integrable UIR in Subsection \ref{IQSQUIR}. Then in Subsection \ref{CartanIQ} we implement this method  with the restriction to the coset issued from the Cartan decomposition of the group, and with the introduction of a weight function defined on a certain submanifold of the Cartan symmetric space.
We apply in Section \ref{SU11cIQ} the above formalism to our specific model of the group SU$(1,1)$, its discrete series, and the unit disk corresponding to the Cartan symmetric space, and we examine the outcomes when a particular  family of weights is considered. 
We then proceed in  Section \ref{permissues} with the quantization of functions on the AdS phase space (i.e. classical observables) and study the dependence of the issue on the choice of the weight function, noticing  that for the most basic ones there is no or trivial dependence.
Section \ref{semclass} is devoted to the lower symbols of operators issued from our covariant integral quantization. This amounts to give semi-classical portraits, in general more regular, of the original function, together with a probabilistic interpretation when the weight function is suitably selected.
In Section \ref{conclu} we conclude by commenting some aspects of our work amenable to further interesting developments. In Section \ref{jacobiint} are given some useful integral formulae involving Jacobi polynomials. 

Most of our approaches should be justified on a mathematical level with regard to involved functions. Nevertheless, they are written here with implicit assumption on their validity on appropriate space of functions (or distributions). Also, throughout the text  we use for convenience  the shortened  notation $f(z)$ in place of $f(z,\bar z)$ for $z\in \C$, at the difference of Berezin in \cite{berezin74}

\section{Geometry of the unit disk and its symmetry}
\label{Dgeom}
\subsection{The unit disk as a K\"{a}hlerian manifold}
\label{Dkaehler}
 The  unit disk
 \begin{equation}
\label{undisk}
\mathcal{D} \deq \{z  \in \C\, , \, \vert z \vert < 1\}
\end{equation}
is one of the 4 two-dimensionial  K\"{a}hlerian manifolds \cite{doubrovine82,perelomov86}, the other ones being respectively the complex plane $\C$, the sphere $\mathbb{S}^2$, or equivalently the projective complex line $\mathbb{CP}^1$,  and the torus $\C/\Z_2 \sim \mathbb{S}^1 \times \mathbb{S}^1$. It is equipped of the  (Poincar\'e) metric
  \begin{equation}
\label{poincrmetric}
\ud s^2 =  \frac{\ud z\, \ud\bar z}{(1-\vert z \vert^2)^2}\, .
\end{equation}
 The corresponding surface  element is given by the two-form:
\begin{equation}
\label{surfelpoinc}
\Omega= \frac{i}{2}\,  \frac{\ud z \wedge \ud\bar z}{(1- \vert z \vert^2)^2}= \frac{\ud(\Re z)\, \ud(\Im z)}{(1- \vert z \vert^2)^2}\equiv \frac{\ud^2z}{(1- \vert z \vert^2)^2} \equiv \mu(\ud^2z)\, .
\end{equation}
 These quantities  are both  issued from a   \emph{K\"{a}hlerian potential} $\mathcal{K}_{\mathcal{D}}$:
\begin{align}
\label{kaelpo1}\mathcal{K}_{\mathcal{D}}(z,\bar z) &\deq \pi^{-1} ( 1- \vert z \vert^2)^{-2} \, , \\
\label{kaelpo2} \ud s^2 &= \frac{1}{2}\,\frac{\partial^2}{\partial z\, \partial \bar z}\, \ln{\mathcal{K}_{\mathcal{D}}(z,\bar z)}\, dz\, d\bar z\, , \\
\label{kaelpo3} \mu(\ud^2z) &=\frac{i}{4}\,  \frac{\partial^2}{\partial z\, \partial \bar z}\, \ln{\mathcal{K}_{\mathcal{D}}(z,\bar z)}\, dz\wedge d\bar z\, .
\end{align}

\subsection{The unit disk as a coset of SU$(1,1)$}
\label{Dcoset}
Let us start by recalling the essential definitions and notations for the simple Lie group SU$(1,1)$ and its Lie algebra.
  \begin{equation}
\label{groupSU11}
\mathrm{SU}(1,1) = \left\{ g = \begin{pmatrix}
 \alpha     &  \beta  \\
\bar \beta      &  \bar \alpha
\end{pmatrix}  \, \ \alpha, \beta \in \C\, , \ \det g= \vert \alpha\vert^2 - \vert \beta\vert^2 = 1\right\}
\end{equation}
The three basis elements  of the Lie algebra $\mathfrak{su}(1,1)$ are chosen as
\begin{equation}
\label{gensu11}
N_0=\frac{1}{2} \begin{pmatrix}
    \ii &  0  \\
    0  &  -\ii
\end{pmatrix}= \frac{\ii \sigma_3}{2}\, , \ N_1=\frac{1}{2} \begin{pmatrix}
    0  &  1  \\
    1  &  0
\end{pmatrix} =\frac{\sigma_1}{2}\, , \  N_2=\frac{1}{2} \begin{pmatrix}
    0  &  \ii  \\
    -\ii  &  0
\end{pmatrix} =-\frac{\sigma_2}{2}\, ,
\end{equation}
with the commutation relations
\begin{equation}
\label{crsu11}
[N_0,N_1]= N_2\, , \quad [N_0,N_2]= -N_1\, , \quad [N_1,N_2]= -N_0\,.
\end{equation}
 The Cartan factorization of SU$(1,1)$ is associated with the (Cartan) involution
 \begin{equation}
\label{cartaninv}
i_{ph}\, : \, g \mapsto (g^{\dag})^{-1}\, .
\end{equation}
The maximal compact  subgroup $H=$U$(1)$ is determined by $i_{ph}(g)= g$ whereas the condition $i_{ph}(g)= g^{-1}$ selects the subset $P$ of Hermitian matrices in  SU$(1,1)$. The factorization SU$(1,1)= PH$ reads explicitly
\begin{equation}
\label{cartanSU11}
\mathrm{SU}(1,1) \ni g = \begin{pmatrix}
 \alpha     &  \beta  \\
\bar \beta      &  \bar \alpha
\end{pmatrix}= p(z)\, h(\theta)\, ,
\end{equation}
with
\begin{equation}
\label{cartanSU11p}
 p(z) = \begin{pmatrix}
   \delta   &   \delta z \\
  \delta \bar z    &  \delta
\end{pmatrix} \,,  \quad  z = \beta \bar{\alpha}^{-1}\, , \quad \delta = \vert \alpha\vert=(1 - \vert z \vert^2)^{-1/2}\, ,
\end{equation}
and
\begin{equation}
\label{cartanSU11h}
 h(\theta) = \begin{pmatrix}
e^{i\theta/2}   &  0 \\
  0   &  e^{-i\theta/2}
\end{pmatrix} \,,  \quad \, \theta = 2\, \arg \alpha\, , \quad 0 \leq \theta < 4 \pi\, .
\end{equation}
 The  \emph{bundle section} $\mathcal{D} \in z  \mapsto p(z) \in  P$ gives the unit disk $\mathcal{D} $
a symmetric space realization identified as  the coset space SU$(1,1)/H$. Note that
\begin{equation}
\label{proppz}
p^2 = g g^{\dag}\, \qquad (p(z))^{-1}= p(-z)\, .
\end{equation}
The Haar measure (see for instance \cite{vilenkin68})  on the unimodular group SU$(1,1)$ from Cartan decomposition
\begin{equation}
\label{HaarSU11}
\ud_{\texttt{haar}} (g) = \frac{1}{8\pi^2}\,\frac{\ud^2 z}{\left( 1- \vert z\vert^2\right)^2}\, \ud\theta\,.
\end{equation}
It is normalized for the angular  parts.

The  Cartan factorization allows to  make SU$(1,1)$ act on $\mathcal{D} $ through a left action on the set of matrices $p( z)$
\begin{equation}
\label{leftactSU11}
g : p(z) \mapsto p(z^{\prime}) \quad \mbox{with} \quad g\, p(z) = p(z^{\prime})\, h'\, ,\quad  
\end{equation}
where  $ z^{\prime}\equiv g\cdot z $ is given by the map
 \begin{align}
\label{homSU11a}
\mathcal{D} \ni z \mapsto z^{\prime}&= (\alpha \, z + \beta)\, (\bar\beta \, z + \bar \alpha)^{-1}  \in \mathcal{D}\\
\label{homSU11b} &\Leftrightarrow z = (\bar\alpha \, z^{\prime} -\beta)\, (-\bar\beta \, z^{\prime} +  \alpha)^{-1} =  g^{-1}\cdot z^{\prime}\, , 
\end{align}
and $h^{\prime}$ is the following element in U$(1)$,
\begin{equation}
\label{hpdef}
h^{\prime}= \begin{pmatrix}
\frac{\beta \bar z + \alpha}{\vert\beta \bar z + \alpha \vert}      &  0  \\
    0  &  \frac{\bar\beta z + \bar \alpha}{\vert\bar\beta z + \bar\alpha \vert} 
\end{pmatrix}\, .
\end{equation}
Note the formula used  to derive the above relations,
\begin{equation}
\label{deltadelta}
\delta(g\cdot z)= \delta(z)\,\vert\bar\beta z + \bar\alpha \vert^{-1}\, ,
\end{equation}
and also the action of U$(1)$ on the disk, 
\begin{equation}
\label{rotdisk}
h(\theta)\,p(z)\,h(-\theta)= p(h(\theta)\cdot z)= p\left(e^{\ii \theta \,z}\right)\, . 
\end{equation}
Hence the unit disk $\mathcal{D}$ is invariant under the  transformations \eqref{homSU11a} of the \emph{homographic} or M\"{o}bius type. 
Note  that SU$(1,1)$ leaves invariant  the boundary $\mathbb{S}^1 \simeq U(1)$ of $\mathcal{D}$ under the transformation (\ref{homSU11a}).

The invariance of $\mathcal{D}$ under (\ref{homSU11a})  also holds for metric quantities  issued from the invariant K\"{a}hlerian potential $\mathcal{K}_{\mathcal{D}}$:
\begin{align}
\label{kaehinv1}
\mathcal{K}_{\mathcal{D}}(z,\bar z) &= \pi^{-1} ( 1- \vert z \vert^2)^{-2} = \pi^{-1} ( 1- \vert z^{\prime} \vert^2)^{-2}\, , \\
\label{kaehinv2} \ud s^2 &= \frac{\ud z\, \ud\bar z}{(1-\vert z \vert^2)^2}= \frac{\ud z^{\prime}\, \ud\bar z^{\prime}}{(1-\vert z^{\prime} \vert^2)^2}\, , \\
\label{kaehinv3} \mu(\ud^2z) &= \frac{\ud(\Re z)\, \ud(\Im z)}{(1- \vert z \vert^2)^2} = \frac{\ud(\Re z^{\prime})\, \ud(\Im z^{\prime})}{(1- \vert z^{\prime} \vert^2)^2}\, .
\end{align}

\subsection{The unit disk as an AdS phase space}
\label{Dphasespace}

Since the unit disk is K\"{a}hlerian, it is symplectic and so can be given a phase space structure and interpretation. Firstly, the form \eqref{surfelpoinc} determines the Poisson bracket
\begin{equation}
\label{poissonb}
\{f,g\}= \frac{1}{2\ii}\,(1-\vert z \vert^2)^2\, \left(\frac{\partial f}{\partial z}\frac{\partial g}{\partial \bar z} - \frac{\partial f}{\partial \bar z} \frac{\partial g}{\partial z}\right)\, .
\end{equation}
Now, there are $3$ basic observables generating the SU$(1,1)$ symmetry on this classical level:
\begin{equation}
\label{3classobs}
\mathcal{D}\ni z \mapsto \ k_0(z)= \frac{1+\vert z\vert^2}{1- \vert z\vert^2}\, , \ k_1(z)= \frac{1}{\ii} \frac{z-\bar z }{1- \vert z\vert^2}\, , \ k_2(z)= \frac{z+\bar z}{1- \vert z\vert^2}\, .
\end{equation}
They are not independent since 
\begin{equation}
\label{uphyp}
k_0^2 -k_1^2 -k_2^2 = 1\, ,
\end{equation}
 i.e., the $3$-vector $(k_0,k_1,k_2)$ points to the upper sheet $\mathcal{H}_+$ of the two-sheeted hyperboloid in $\R^3$ which is described by \eqref{uphyp}, and whose the stereographic projection through \eqref{3classobs} is the open unit disk. This projection reads
 \begin{equation}
\label{hypdisk}
\mathcal{H}_+ \ni (k_0,k_1,k_2) \mapsto z= \frac{k_2 + \ii k_1}{1 + k_0}\equiv \sqrt{\frac{k_0 -1}{k_0 + 1}} e^{\ii \arg z}\,. 
\end{equation} 
They obey the Poisson commutation rules
\begin{equation}
\label{Poissoncr}
\{k_0,k_1\}= k_2\, , \quad \{k_0,k_2\}= - k_1\, , \quad  \{k_1,k_2\}= - k_0\, ,
\end{equation}
which are consistent with \eqref{crsu11}. As is expected,  the two combinations
\begin{equation}
\label{combkk}
k_+ = k_2 -\ii k_1 = \frac{2\bar{z}}{1-\vert z \vert^2}\, , \quad k_- = k_2  + \ii k_1 = \frac{2z}{1-\vert z \vert^2}\, ,
\end{equation}
are to play an important role as well.

Under the action of $g= \begin{pmatrix}
   \alpha   & \beta   \\
  \bar{\beta}    &  \bar{\alpha}
\end{pmatrix}\in \text{SU}(1,1)$, functions $k_0$ and $k_{\pm}$ transform as
\begin{align}
\label{k0g}
  k_0^{\prime}(z)=k_0\left(g^{-1}\cdot z\right)  &= \left(\vert \alpha\vert^2 + \vert \beta\vert^2\right)\,k_0(z) - 2\Re\left(\alpha\beta k_{+}(z)\right) \, ,  \\
 \label{k+g} k^{\prime}_{+}(z)=  k_{+}\left(g^{-1}\cdot z\right) &= -2\alpha \bar{\beta}\,k_0(z) + \alpha^2\, k_{+}(z) + \bar{\beta}^2\, k_{-}(z)  \, ,\\
 \label{k-g} k^{\prime}_{-}(z)  =  k_{-}\left(g^{-1}\cdot z\right) &= -2\bar{\alpha} \beta\,k_0(z)  + \bar{\alpha}^2\, k_{-}(z)+ \beta^2\, k_{+}(z) =  \overline{k_{+}\left(g^{-1}\cdot z\right)} \, .  
\end{align}
Equivalently,
\begin{equation}
\label{gkg}  \begin{pmatrix}
  k^{\prime}_{-}    &  k^{\prime}_0  \\
   k^{\prime}_0   &  k^{\prime}_{+}
\end{pmatrix} =
g^{-1} \begin{pmatrix}
  k_{-}    &  k_0  \\
   k_0   &  k_{+}
\end{pmatrix} \left(g^{-1}\right)^{t}\,.
\end{equation}
This transform can be viewed as the (co-adjoint action) of SU$(1,1)$ on the co-adjoint orbit identified with \eqref{uphyp}. Introducing the left action of SU$(1,1)$ on functions $f(z)$
as
\begin{equation}
\label{leftaction}
(\mathfrak{U}(g)f)(z):= f\left(g^{-1}\cdot z\right)\, , 
\end{equation}
we write the above transformations in terms of matrix elements as
\begin{equation}
\label{Umatekab}
(\mathfrak{U}(g)k_a)(z)= \sum_b [\mathfrak{U}(g)]_{ba} k_b(z)\, , \quad a=0, 1,2\ \mbox{or} \ a =0, \pm \, . 
\end{equation}
The following particularisation of these formulae to $g=p(-z^{\prime})$ will be useful for Section \ref{semclass}:
 \begin{align}
\label{k0p}
  k_0(p(z^{\prime})\cdot z)&= k_0(z^{\prime})\,k_0(z) + \Re\left(k_-(z^{\prime}) \,k_{+}(z)\right) \, ,  \\
 \label{k+p} k_{+}(p(z^{\prime})\cdot z) &=k_+(z^{\prime})\,k_0(z) + \left(1-\vert z^{\prime}\vert^2\right)\left(k_{+}(z) + {\bar{z^{\prime}}}^2\, k_{-}(z) \right) \, ,\\  \label{k-p} k_{-}(p(z^{\prime})\cdot z) &= \overline{k_{+}(p(z^{\prime})\cdot z) } \, .  
\end{align}

\subsection{AdS space-time as a left coset of  SU$(1,1)$}
\label{AdScoset}

 \begin{equation}
\label{factSU11}
g =   h(\theta)\, s(u)\, l(v)\, , \quad \theta \in [0,2\pi),\, \,u,v\in \R.
\end{equation}
 Whereas  the first factor
 \begin{equation}
\label{htheta}
h(\theta) = \epsilon \begin{pmatrix}
  e^{\ii \theta/2}    &  0  \\
   0   &  e^{-\ii \theta/2}
\end{pmatrix}\, , \quad  \quad \epsilon = \pm I_2 \, ,
\end{equation}
 belongs to U$(1)$, the maximal  compact subgroup, with $\epsilon$ belonging to the center of SU$(1,1)$ isomorphic to $\Z_2$,  and the others are of non-compact hyperbolic type:
\begin{equation}
\label{hypsulv}
s(u)= \begin{pmatrix}
 \cosh \frac{u}{2}     &    \sinh \frac{u}{2}\\
    \sinh \frac{u}{2}   &   \cosh \frac{u}{2}
\end{pmatrix}\, , \ l(v)= \begin{pmatrix}
 \cosh \frac{v}{2}    &   \ii\,  \sinh \frac{v}{2}\\
 -\ii\,    \sinh \frac{v}{2}   &   \cosh \frac{v}{2}
\end{pmatrix}\,, \quad u, \, v \in \R\,  ,
\end{equation}
and belong to subgroups  isomorphic to $\R$.
Their respective   generators $N_a$, $a = 0, 1, 2$ are precisely those introduced in \eqref{gensu11},
\begin{equation}
\label{genexpsu11}
h(\theta) = \epsilon e^{\theta \, N_0}\, , \quad s(u) = e^{u \, N_1} \, , \quad l(v) = e^{v \, N_2}\, .
\end{equation}
The factorisation \eqref{factSU11} is associated with the  group involution
\begin{equation}
\label{ stinv}
i_j : g \mapsto g^t
\end{equation}
where the superscript $t$ denotes transposition. Indeed,
\begin{equation}
\label{invhsl}
h^t = h\, , \quad s^t = s\, , \quad l^t= l^{-1} \, .
\end{equation}
Now by considering
\begin{equation}
\label{defj}
j(\theta, u)= h(\theta)\, s(u)= \begin{pmatrix}
e^{\ii \theta/2}  \cosh \frac{u}{2}     &  e^{\ii \theta/2}   \sinh \frac{u}{2}\\
e^{-\ii \theta/2}     \sinh \frac{u}{2}   &  e^{-\ii \theta/2}  \cosh \frac{u}{2}
\end{pmatrix}\, ,
\end{equation}
we have
\begin{equation}
\label{jjt}
jj^t= gg^t= \begin{pmatrix}
 \alpha^2 + \beta^2     &   2\Re(\alpha \bar \beta) \\
2\Re(\alpha \bar \beta)      &  \bar\alpha^2 + \bar\beta^2
\end{pmatrix} = \begin{pmatrix}
e^{\ii \theta}  \cosh u     &     \sinh u\\
     \sinh u   &  e^{-\ii \theta}  \cosh u
\end{pmatrix}\, .
\end{equation}
The parameters $(\theta, u)$ form a system of global coordinates for the $1+1$-Anti de Sitter space-time visualized as the one-sheeted hyperboloid $\eta_{ab}y^a y^b= \kappa^{-2}$  in $\R^3$ with metric $(\eta^{ab}) = \mathrm{diag} (+,+,-)$, $a,b= 2,0, 1$,  and with curvature $\kappa$,
\begin{equation}
\label{adshyp}
y^2= \kappa^{-1}\cosh u \cos\theta\,, \quad y^0= \kappa^{-1}\cosh u \sin\theta\,, \quad y^1= \kappa^{-1}\sinh u\,,
\end{equation}
or expressed in terms of the element of SU$(1,1)$,
\begin{equation}
\label{Gamy}
jj^t= \begin{pmatrix}
\kappa y_+    &   \kappa y^1\\
\kappa y^1     &  \kappa y_-
\end{pmatrix} \equiv \Gamma(y)\, , \quad y_{\pm} = y^2 \pm \ii y^0\, , \quad \det \Gamma(y) =  \eta_{ab}y^a y^b= \kappa^{-2}\, .
\end{equation}
The factorisation $g=jl$, which means on the group level SU$(1,1) = $ AdS$\,\times\,$Lorentz, allows to view the AdS space-time as the left coset SU$(1,1)/$L, with L $= \{l(v)\,, \,  v\in \R\}\sim$ SO$_0(1,1)$ is the orthochronous Lorentz subgroup.
SU$(1,1)$ acts on the set of matrices $\Gamma(y)$  and this action is induced from its left action  on the set of matrices $j$
\begin{equation}
\label{actgj}
g\, :\, j\mapsto j^{\prime}\, , \  gj= j^{\prime}l^{\prime}\, \ \Leftrightarrow \Gamma(y^{\prime}) = j^{\prime}{j^{\prime}}^t= gjj^tg^t = g\Gamma(y)g^t\, .
\end{equation}
In this interpretation, SU$(1,1)$ acts as  the double covering of the  actual AdS group SO$_0(2,1)=$ SU$(1,1)/\Z_2$, and we see that $N_0$ generates the ``translations in time'' corresponding to $U(1)$, $N_1$ generates the ``translations in space'' corresponding to the subgroup SO$_0(1,1)$, and $N_{2}$ generates the Lorentz transformations corresponding to the other SO$_0(1,1) =$\,L.

It is instructive to describe how the three basic observables $k_a$, $a=0,1,2$ (or $k_0$, $k_{\pm}$),  transform under the action of three subgroups with respective generators $N_a$, $a=0,1,2$.
\begin{align}
\label{N0k}
 h(\theta)   &\left\lbrace\begin{array}{ccl}
   k_0   &  \mapsto & k_0  \\
    k_{\pm}  &   \mapsto & e^{\pm\ii \theta}k_{\pm} 
\end{array} \right. \, ,   \\
 \label{N1k}   s(u)   &\left\lbrace\begin{array}{ccl}
   k_0   &  \mapsto & \cosh u \,k_0 -\sinh u\,k_2 \\
    k_1  &   \mapsto & k_1 \\
    k_2   &  \mapsto & -\sinh u \,k_0 +\cosh u\,k_2
     \end{array} \right. \, ,   \\
\label{N2k}    l(v)   &\left\lbrace\begin{array}{ccl}
   k_0   &  \mapsto & \cosh v \,k_0 -\sinh v\,k_1 \\
    k_1  &  \mapsto & -\sinh v \,k_0 +\cosh u\,k_1 \\
      k_2  &   \mapsto & k_2 
     \end{array} \right. \, . 
\end{align}

\section{SU$(1,1)$  representation(s)}
\label{SU11rep}

\subsection{SU$(1,1)$ unitary irreducible representation(s) (discrete series)}
\label{discreteseries}
For a given $\eta > 1/2$, consider the Fock-Bargmann Hilbert space $\mathcal{FB}_{\eta}$ of all analytic functions $f(z)$ on $\mathcal{D}$ that are square integrable with respect to the scalar product
\begin{equation}
\label{FBsu11inprod}
\lg f_1|f_2\rg = \frac{2 \eta -1}{2 \pi} \, \int_{\mathcal{D}} \overline{f_1(z)}\, f_2(z)\, (1-\vert z \vert^2)^{2 \eta -2}\, \ud^2z\, .
\end{equation}
An orthonormal basis is made of powers of $z$  suitably normalized:
\begin{equation}
\label{orthbasdisk2}
e_{n} (z) \equiv \,  \sqrt{\frac{(2 \eta)_n}{n!}}\,  z^n \quad   \mathrm{with}\quad  n \in \N,
\end{equation}
where $(2 \eta)_n:=\Gamma(2 \eta +n)/\Gamma(2 \eta)$ is the Pochhammer symbol. For $\eta = 1, 3/2, 2, 5/2, \dotsc,$ one defines the UIR $g =\begin{pmatrix}
 \alpha     &  \beta  \\
  \bar \beta    & \bar \alpha
\end{pmatrix} \mapsto U^{\eta}(g)$ of $SU(1,1)$ on  $\mathcal{FB}_{\eta}$ by:
 \begin{equation}
\label{UIRDSsu11}
\mathcal{FB}_{\eta} \ni f(z) \mapsto \left(U^{\eta}(g)\, f\right)(z) = (-\bar \beta \, z + \alpha)^{-2 \eta}\, f \left( \frac{\bar \alpha z - \beta}{- \bar \beta z + \alpha}\right)\, .
\end{equation}
In particular for $g=p(z^{\prime})$ we have
\begin{equation}
\label{UIRDSsu11}
 \left(U^{\eta}(p(z^{\prime}))\, f\right)(z)
 = (1-|z^{\prime}|^2)^{\eta}\, (1-z\,\bar z^{\prime})^{-2\eta}\, f \left( \frac{ z - z^{\prime}}{1- z\,\bar z^{\prime}}\right)\, .
\end{equation}
This countable set of representations constitutes the ``almost complete''  holomorphic discrete series of representations of SU$(1,1)$.  ``Almost complete'' because the lowest one, $\eta = 1/2$, requires a special treatment due to the non existence of the inner product  (\ref{FBsu11inprod}) in this case.
Had we considered the continuous set $\eta \in [1/2, + \infty)$, we would have been  led to involve the universal covering of $SU(1,1)$

The matrix elements of the operator $U^{\eta}(g)$ with respect to the orthonormal basis (\ref{orthbasdisk2}) are given  in terms of hypergeometric polynomials by:
\begin{align}
\label{mateldiscsu11}
\nonumber U^{\eta}_{nn^{\prime}}(g) &= \lg e_n | U^{\eta}(g) | e_{n^{\prime}}\rg = \left( \frac{n_>!\, \Gamma(2 \eta + n_>) }{n_<!\, \Gamma(2 \eta + n_<)} \right)^{1/2}\, {\alpha}^{-2 \eta - n_>}\, \bar{\alpha}^{n_<} \times \\
&\times\frac{(\gamma(\beta,\bar \beta))^{n_> -n_<}}{(n_> - n_<)!}\,  {}_2F_1\left(-n_<\, ,\, n_> + 2\eta\, ;\,  n_> - n_< +1\, ;\, \frac{\vert \beta\vert^2}{\vert \alpha\vert^2} \right)\, ,
\end{align}
where
\begin{equation*}
\gamma(\beta,\bar \beta) = \left\lbrace \begin{array}{cc}
   - \beta  &  n_> =  n^{\prime} \\
   \bar \beta    &   n_> = n
\end{array}\right.\, , \quad n_{\substack{
>\\
<}}  = \left\lbrace \begin{array}{c}
    \max      \\
       \min
\end{array}\right.\, (n,n^{\prime}) \geq 0\, .
\end{equation*}
Taking into account the well-known relation between the hypergeometric functions and the Jacobi \cite{magnus66} polynomials
\[
P^{(\mu ,\nu)}_{n}\left(x\right)=
\left(\begin{array}{c} n+\mu \\
n\end{array}\right)\,
{}_2F_1\left(-n\, ,\, n + \mu+\nu+1\, ;\,  \mu +1\, ;\, \frac{1-x}{2} \right)
\]
and the parametrization \eqref{cartanSU11p}, this expression is alternatively given  in terms of Jacobi polynomials as:
\begin{align}
\label{mateldiscsu11Jac}
\nonumber U^{\eta}_{nn^{\prime}}(g) &= \left( \frac{n_<!\, \Gamma(2 \eta + n_>) }{n_>!\, \Gamma(2 \eta + n_<)} \right)^{1/2}\, \alpha^{-2 \eta - n_>}\, \bar{\alpha}^{n_<} \times \\
&\times (\gamma(\beta,\bar \beta))^{n_> -n_<}\, P^{(n_>-n_<\, ,\, 2 \eta -1)}_{n_<}\left( 1-2\vert z\vert^2 \right)\,, \quad z = \beta \bar{\alpha}^{-1}\, .
\end{align}
Note the  diagonal elements,
\begin{equation}
\label{ matdiag}
U^{\eta}_{nn}(g)  =  {\alpha}^{-2 \eta - n}\, \bar{\alpha}^{n} \, {}_2F_1\left(-n\, ,\, n + 2\eta\, ;\, 1\, ;\, \frac{\vert \beta\vert^2}{\vert \alpha\vert^2} \right)= \alpha^{-2 \eta}\, \left(\frac{\bar{\alpha}}{\alpha}\right)^nP^{(0\, ,\, 2 \eta -1)}_{n}\left( 1-2\vert z\vert^2 \right)\, .
\end{equation}
For the elements  $g=h(\theta)$ in U$(1)$, we have
\begin{equation}
\label{Uhtheta}
U^{\eta}_{nn^{\prime}}(h(\theta)) =  \delta_{n n^{\prime}}\, e^{-\ii(\eta + n) \theta}\, ,
\end{equation}
whereas for the elements  $g=p(z)$ in $P$,
\begin{align}
\label{Upz}
\nonumber   U^{\eta}_{nn^{\prime}}(p(z)) &= \left( \frac{n_>!\, \Gamma(2 \eta + n_>) }{n_<!\, \Gamma(2 \eta + n_<)} \right)^{1/2}  \left(1-\vert z \vert^2\right)^{\eta} \, \frac{\vert z \vert^{n_> -n_<}}{(n_> - n_<)!}\, e^{\ii(n^{\prime}-n)\phi}\times \\
&\times  (\mathrm{sgn}(n-n^{\prime}))^{n-n^{\prime}} {}_2F_1\left(-n_<\, ,\right.\left. n_> + 2\eta\, ;n_> - n_< +1\, ;\, \vert z\vert^2 \right)\\
\nonumber  & = \left( \frac{n_<!\, \Gamma(2 \eta + n_>) }{n_>!\, \Gamma(2 \eta + n_<)} \right)^{1/2}  \left(1-\vert z \vert^2\right)^{\eta} \, \vert z \vert^{n_> -n_<}\, e^{\ii(n^{\prime}-n)\phi}\times \\
& \times  (\mathrm{sgn}(n-n^{\prime}))^{n-n^{\prime}}\, P^{(n_> - n_<\, ,\, 2 \eta -1)}_{n_<}\left( 1-2\vert z\vert^2 \right)
\end{align}
with $z = \vert z \vert e^{\ii\phi}$, and if $n=n^{\prime}$,
\begin{equation}
\label{matdiagpz}
U^{\eta}_{nn}(p(z))  =  (1-\vert z \vert^2)^{\eta} \, {}_2F_1\left(-n\, ,n + 2\eta\, ; 1\, ;\, \vert z\vert^2 \right)= \left(1-\vert z \vert^2\right)^{\eta}\, P^{(0\, ,\, 2 \eta -1)}_{n}\left(1-2\vert z\vert^2\right)\, .
\end{equation}

\subsection{Orthogonality relations and trace formulae}

Since the functions \eqref{mateldiscsu11Jac} are matrix elements of  operators $U^{\eta}$ in the discrete series for $\eta> 1/2$, one of their fundamental properties is displayed by their orthogonality relations:
\begin{equation}
\label{orthogrel}
\int_{\mathrm{SU}(1,1)}\ud_{\mathrm{haar}}(g)\, U^{\eta}_{mm^{\prime}}(g)\, \overline{U^{\eta}_{nn^{\prime}}(g)}= d_{\eta}\,\delta_{mn}\delta_{m^{\prime}n^{\prime}}\, ,
\end{equation}
where $d_{\eta}=2\pi/(2\eta-1)$ is the (formal) dimension of the representation $U^{\eta}$.

From the extension of the generating function of Jacobi polynomials \cite{magnus66}  
\begin{equation}
\label{genjac}
\sum_{n=0}^{\infty} P^{(\mu\, ,\, \nu)}_{n}\left(x\right) \,t^n= \frac{2^{\mu + \nu}}{R}\, (1-t+R)^{-\mu}\, (1+t+R)^{-\nu}\, , \quad \vert t \vert < 1\, , \ R = (1-2tx + t^2)^{1/2}\, , 
\end{equation}
to its ``forbidden'' limit $\vert t \vert = 1$, 
we infer the trace of the operator  $U^{\eta}(g)$ for a general $g= \begin{pmatrix}
  \alpha    &  \beta  \\
   \bar\beta   &  \bar\alpha
\end{pmatrix} \in \mathrm{SU}(1,1)$, 
\begin{equation}
\label{traceUng}
\mathrm{tr}\,\left(U^{\eta}(g)\right)= \frac{1}{2}\,\left((\Re\alpha)^2-1\right)^{-1/2}\, \left[(\Re\alpha)^2 + \left((\Re\alpha)^2-1\right)^{1/2}\right]^{1-2\eta}\, , 
\end{equation}
and its restriction to $p(z)$, 
\begin{equation}
\label{traceUn}
\mathrm{tr}\,\left(U^{\eta}(p(z))\right)=\frac{1}{2\vert z\vert}\,(1-\vert z \vert^2)^{\eta}(1+\vert z\vert)^{1-2\eta}\,. 
\end{equation}
These expressions are singular for $\alpha = 1$ and  $z= 0$ respectively, since these values correspond to the identity operator. 

Another trace formula will play an important role in the sequel. It involves the parity operator defined by
\begin{equation}
\label{paritydef}
\sfP := \sum_{n=0}^{\infty}(-1)^n |e_n\rg\lg e_n|\,.
\end{equation}
\begin{equation}
\label{traceUngP}
\mathrm{tr}\,\left(\sfP\, U^{\eta}(g)\right)= \frac{1}{2}\,\left((1-\Im\alpha)^2\right)^{-1/2}\, \left[(\Im\alpha)^2 + \left((1-\Im\alpha)^2\right)^{1/2}\right]^{1-2\eta}\, , 
\end{equation}
and its restriction to $p(z)$, 
\begin{equation}
\label{traceUnP}
\mathrm{tr}\,\left(\sfP \,U^{\eta}(p(z))\right)=\frac{1}{2}\,. 
\end{equation}
For $z=0$ this gives a trace formula for the parity operator
\begin{equation}
\label{traceP}
\mathrm{tr}\,\sfP = \sum_{k=0}^{\infty}(-1)^k= \frac{1}{2}\, , 
\end{equation}
which can be legitimated   by adopting the Abel summation of divergent series.

Finally, the following formula (related to the existence of the so-called inversion in Cartan symmetric domains) will be used in this paper.
\begin{equation}
\label{Ppz}
\sfP\,U^{\eta}(p(z))\, \sfP= U^{\eta}(p(-z))\, .
\end{equation}


\subsection{Corresponding representation of the Lie algebra $\mathfrak{su}(1,1)$}
\label{su11}
 The respective self-adjoint representatives of $N_0$, $N_1$, and $N_2$, defined in \eqref{gensu11}, under the UIR (\ref{UIRDSsu11}), defined generically  as $\left.\ii \, \partial/\partial t \, U^{\eta}(g(t))\right|_{t=0}$, are the following differential operators  on the Fock-Bargmann space $\mathcal{FB}_{\eta}$:
\begin{subequations}\label{uirepsu11}
\begin{align}
\label{uirepsu110}
N_0  &\mapsto K_0 = z \,\frac{\ud}{\ud z}  + \eta\, , \\
  \label{uirepsu111}    N_1 &  \mapsto K_1 = - \frac{\ii}{2}(1- z^2)\, \frac{\ud}{\ud z} + \ii \eta z\, , \\
 \label{uirepsu112}         N_2 &  \mapsto K_2 =  \frac{1}{2}(1 + z^2)\, \frac{\ud}{\ud z} +  \eta z\, ,
\end{align}
\end{subequations}
They obey the commutation rules,
\begin{equation}
\label{comreluirsu11}
[K_0, K_1] = \ii\, K_2\, , \quad [K_0,K_2] = -\ii\, K_1\, , \quad [K_1,K_2]  = -\ii\,  K_0\, .
\end{equation}
The elements of the orthonormal basis (\ref{orthbasdisk2})  are eigenvectors of the compact generator $K_0$ with equally spaced  eigenvalues:
\begin{equation}
\label{spectrmK0}
K_0 \, |e_n\rg = (\eta + n )\, |e_n\rg\, .
\end{equation}
The particular element  $|e_0\rg$ of the basis is a \emph{lowest weight} state or ``vacuum''  for the representations $U^{\eta}$.
 Indeed, introduce  the two operators with their commutation relation:
\begin{equation}
\label{raislowsu11}
K_{\pm} = \mp \ii\, (K_1 \pm \ii K_2) = K_2 \mp \ii K_1\, , \quad [K_+, K_-] = -2 K_0\, .
\end{equation}
 As differential operators, they read as: $K_+ = z^2\, \ud/\ud z + 2 \eta z$, $K_- = \ud/\ud z$.  Adjoint of each other, they are raising and lowering operators respectively:
\begin{equation}
\label{raislowstaFBsu11}
 K_+\, |e_n\rg = \sqrt{(n+1)(2 \eta + n)}\, |e_{n+1}\rg\, ,\quad \quad K_-\, |e_n\rg  = \sqrt{n \, (2 \eta + n-1)}\, |e_{n-1}\rg\, ,
\end{equation}
and we check $K_- \, |e_0\rg = 0$.
States  $ |e_n\rg$ are themselves obtained by successive ladder actions on the lowest state as follows:
\begin{equation}
\label{laddersu11}
|e_n\rg  = \sqrt{\frac{\Gamma(2\eta)}{\Gamma(2 \eta + n) \, n!}}\, (K_+)^n\, |e_0\rg\, .
\end{equation}
The Casimir operator is defined as
\begin{equation}
\label{casruidiscsu11}
\mathcal{C} \deq  K_1^2 + K_2^2 -K_0^2 = \frac{K_+K_- + K_- K_+}{2}- K_0^2
\end{equation}
This operator is fixed at the value $\mathcal{C} = -\eta (\eta -1)\, I_d$ on the space $\mathcal{FB}_{\eta} $ that carries the UIR $U^{\eta}$.

Finally, we note that, in agreement with the covariance properties \eqref{k0g}, \eqref{k+g}, and \eqref{k-g}, of their respective classical counterparts, we have
\begin{align}
\label{K0g}
  U^{\eta}(g)\,K_0\, U^{\eta}\left(g^{-1}\right) &= \left(\vert \alpha\vert^2 + \vert \beta\vert^2\right)\,K_0- \alpha\beta K_{+} - \bar{\alpha}\bar{\beta} K_{-}\, ,  \\
 \label{K+g} U^{\eta}(g)\,K_{+}\, U^{\eta}\left(g^{-1}\right)  &= -2\alpha \bar{\beta}\,K_0 + \alpha^2\, K_{+} + \bar{\beta}^2\, K_{-}  \, ,
 \\
 \label{K-g} U^{\eta}(g)\,K_{-}\, U^{\eta}\left(g^{-1}\right)  &= -2\bar \alpha \beta\,K_0 + \bar{\alpha}^2\, K_{-} + \beta^2\, K_{+}  \, .
 \end{align}
 Equivalently, with the notations of \eqref{Umatekab}
\begin{equation}
\label{UmateKab}
U^{\eta}(g)\,K_a\, U^{\eta}\left(g^{-1}\right) = \sum_b [\mathfrak{U}(g)]_{ba} K_b(z)\, , \quad a=0, 1,2\ \mbox{or} \ a =0, \pm \, . 
\end{equation}
Like for the Weyl-Heisenberg group, a unitary ``displacement'' operator is built from the generators $K_{\pm}$. It corresponds to map $\mathcal{D} \ni z \mapsto \xi \in \C$ determined by
\begin{equation}
\label{UpzeK}
U^{\eta} (p(\bar z)) = e^{\xi\, K_+ - \bar \xi \, K_-}\equiv D_{\eta}(\xi)\, , \quad \xi= \tanh^{-1}\vert z\vert\,e^{\ii \arg z}\, , 
\end{equation}
which gives $k_0(z)= \cosh 2\vert\xi\vert$  for the observable  introduced in \eqref{3classobs}.

\section{Covariant integral quantizations : General}
\label{covIQ}

\subsection{Covariant integral quantization with  UIR of a group}
\label{CIQUIR}
Here we recall the general principles of this method.
 Let $G$ be a Lie group with left Haar measure $\ud_{\texttt{haar}}(g)$, and let $g\mapsto U(g)$ be a UIR of $G$ in a Hilbert space $\mathcal{H}$.
Let ${\sf M}$ be a bounded operator on $\mathcal{H}$ and let us introduce the family
\begin{equation}
\label{famtransp}
\{ {\sf M}(g):= U(g)\, {\sf M}\, U^{\dag}(g)\, , \, g\in G\}
\end{equation}
of ``displaced'' version of $M$ under the action of the $U(g)$'s.  Suppose that the  operator
\begin{equation}
\label{intgrR}
R:= \int_G  \, {\sf M}(g)\,\ud_{\texttt{haar}}(g) \, ,
\end{equation}
is defined in a weak sense. From the left invariance of $\ud_{\texttt{haar}}(g)$  the operator $R$ commutes with all operators $U(g)$, $g\in G$, and so, from Schur's Lemma, we have the ``resolution'' of the unity up to a constant,
\begin{equation}
\label{resunitG}
R= c_{ {\sf M}}I
\end{equation}
with
\begin{equation}
\label{calcrho}
c_{ {\sf M}} = \int_G  \, \mathrm{tr}\left(\rho_0\, {\sf  M}(g)\right)\, \ud_{\texttt{haar}}(g)\, .
\end{equation}
Here the unit trace positive operator $\rho_0$ is  chosen, if manageable,  in order to make the integral convergent.  Of course, it is possible that no such finite constant exists for the given ${\sf M}$, and worst, it could not exist for any ${\sf M}$ (which is not the case for square integrable representations).
Now, if $c_{ {\sf M}}$ is finite and positive, the true resolution of the identity follows:
\begin{equation}
\label{Resunityrho}
\int_G \, {\sf M}(g) \,\ud \nu(g) = I\,, \quad \ud \nu(g):= \ud_{\texttt{haar}}(g)/c_{ {\sf M}}\, .
\end{equation}

\subsection{Covariant  integral quantization: with square integrable UIR}
\label{IQSQUIR}

Let us consider a  UIR $U$  for which ${\sf M}$ is an ``admissible''  operator, which means that
 \begin{equation}
\label{cgM}
c_{{\sf M}}= \int_G \ud_{\texttt{haar}}(g) \,  \mathrm{tr} \rho_0 {\sf M}(g)
\end{equation}
is finite for a certain $\rho_0$, or more specifically,
for square-integrable UIR $U$  for which ${\sf M}= \rho$ is an admissible density operator,
\begin{equation}
\label{cGrho}
c(\rho)= \int_G \ud_{\texttt{haar}}(g) \, \Vert \rho U(g)\Vert_{\mathcal{HS}}^2   < \infty\, ,
\end{equation}
where $ \Vert  A \Vert_{\mathcal{HS}}= \mathrm{tr}\left(A A^\dag\right)$ is the Hilbert-Schmidt norm.
Then,
the resolution of the identity  is guaranteed with the family:
\begin{equation}
\label{transpM}
{\sf M}(g)= U(g) {\sf M} U^{\dag}(g) \, .
\end{equation}
 This allows the \textit{covariant} integral  quantization of complex-valued functions on the group
 \begin{equation}
\label{GquantMg}
f\mapsto A_f = \int_G\,  {{\sf M} (g)}\,f(g)\, \ud \nu(g)\, .
\end{equation}
Covariance means that
\begin{equation}
\label{covintquant}
U(g) A_f U^{\dag}(g) = A_{U_{\texttt{reg}}(g)f}\, ,
\end{equation}
where
\begin{equation}
\label{regrep}
(U_{\texttt{reg}}(g)f)(g^{\prime}):= f(g^{-1}g^{\prime})
\end{equation}
is the regular representation if $f\in L^2(G,\ud_{\texttt{haar}}(g))$.
Moreover, we get a  generalization of the Berezin or heat kernel transform on $G$ (see for instance \cite{hall06}):
\begin{equation}
\label{genber}
f(g) \mapsto \check{f}(g) := \int_{G}\, \mathrm{tr}(\rho(g)\,\rho(g^{\prime})) \, f(g^{\prime})\,\ud\nu(g^{\prime})\,;
\end{equation}
where the function $\check{f}$ is the \textit{lower} or \textit{covariant symbol} of the operator $A_f$. 
\subsection{Covariant integral quantization through Cartan decomposition: the general case}
\label{CartanIQ}
\subsubsection*{Cartan decomposition: a reminder}
Let $G$ be a connected semi-simple Lie group and $K$ its maximal compact subgroup. Then the homogeneous coset space
  \begin{equation}
\label{symspace}
P= G/K
\end{equation}
is symmetric (i.e., is a smooth manifold whose group of symmetries contains an inversion symmetry about every point), diffeomorphic to a Euclidean space,  and the Cartan decomposition
 \begin{equation}
\label{cartandec}
G=P\,K\Leftrightarrow \ \forall g\in G\  \exists\, p\in P\, , \, k\in K\,, \ g= pk = k p^{\prime}\, , \ p^{\prime}= k^{-1}p k\,,
\end{equation}
holds.
This decomposition exponentiates the Lie algebra decomposition
\begin{equation}
\label{LieCartan}
\mathfrak g = \mathfrak p + \mathfrak k\, ,  \ [\mathfrak k\,, \,\mathfrak k] \subset\mathfrak k\, , \ [\mathfrak k\,, \,\mathfrak p] \subset\mathfrak p\, , \ [\mathfrak p\,, \,\mathfrak p] \subset\mathfrak k\, .
\end{equation}
The  action $g: p\mapsto g\cdot p= p^{\prime}$ of $G$ on $P$ is carried out  through the left action $gp=p^{\prime}k^{\prime}$. Hence, the subgroup $K$ is the stabilizer of a point in $P$.
 The corresponding Cartan involution  is denoted by $\vartheta$:
\begin{equation}
\label{cartaninv}
\vartheta(p) = p^{-1} \ \forall\, p\in P\,, \quad \vartheta(k)  = k \ \forall\, k\in K\, , \quad \vartheta(\mathfrak p)= - \mathfrak p\, , \  \vartheta(\mathfrak k)= \mathfrak k
\end{equation}
 A derived decomposition  involves the abelian subgroup $A= \exp \mathfrak a$ where $\mathfrak a$ is a (Cartan) maximal abelian subalgebra in $\mathfrak p$
\begin{equation}
\label{CartanKAK}
G= KAK\, .
\end{equation}
 Geometrically the image of the subgroup $A$ in $P= G / K$  is a totally geodesic submanifold.
   Since $G$ is unimodular, the Haar measure is left and right invariant and can be factorized as
   \begin{equation}
\label{facthaarG}
\ud_{\texttt{haar}}(g) = \ud\mu_P(p)\,\ud_{\texttt{haar}}(k)
\end{equation}
 with the invariance property for $\ud\mu_P$
 \begin{equation}
\label{invmuP}
\ud\mu_P(kpk^{\prime})= \ud\mu_P(p) \ \forall\,k,k^{\prime} \in K\,.
\end{equation}
Due to \eqref{CartanKAK}, the measure $\ud\mu_P$ can be factorized as
\begin{equation}
\label{muPfact}
\ud\mu_P(p) = \ud_{\texttt{haar}}(k^{\prime})\,\ud\mu_A(a)\,
\ud_{\texttt{haar}}(k^{\prime\prime})\, ,
\end{equation}
where, due to the Euclidean nature of $\mathfrak a$,
\begin{equation}
\label{measA}
\ud\mu_A(a) = \sigma(a) \, \prod_{i}\ud a_i\, ,
\end{equation}
the function $a\mapsto \sigma(a)$ being left and right $K$-invariant
\begin{equation}
\label{invsigma}
\sigma(kak^{\prime}) = \sigma(a)\quad \forall\, k,k^{\prime}\in K\,.
\end{equation}

\subsubsection*{Integral quantization of Cartan symmetric space}
Now,  let $a\mapsto w(a)$ be a  function  on $A$ which is left and right $K$-invariant (it can be complex-valued) and $U$ a UIR of $G$. Suppose that this function allows to define
  \begin{equation}
\label{defMw}
\sfM^{w} := \int_P \ud\mu_P(p) \, w(a(p))\, U(p)\,.
\end{equation}
as an operator bounded (in a weak sense). Introducing its displaced version under the action of $U$
  \begin{equation}
\label{transpMU}
\sfM^{w}(g)= U(g)\,\sfM^{w}\,U^\dag(g)\, ,
\end{equation}
and supposing that the Haar measure on $K$ is normalized, one derives from \eqref{Resunityrho} the resolution of the identity
\begin{equation}
\label{residMw}
\int_P \frac{\ud\mu_P(p)}{C^{w}} \,\sfM^{w}(p) = I\,, \qquad C^{w} = \int_P \ud\mu_P(p)\, \mathrm{tr}\left(\rho_0\,\sfM^{w}(p)\right)\,,
\end{equation}
where the density operator $\rho_0$ has been suitably chosen.

 Effectively, starting from \eqref{Resunityrho} and from the integral representation of $\sfM^{w}$,
\begin{align*}
 I&= \int_G  \frac{\ud_{\texttt{haar}}(g)}{C^{w}}\, {\sf M}^w(g) \\
 &= \int_P \frac{\ud\mu_P(p)}{C^{w}} \,\int_K \ud_{\texttt{haar}}(k)\,\int_P \ud\mu_P(p^{\prime})\,w(a(p^{\prime}))\,U(pk)\, {\sf M}^w(p^{\prime})\,U^\dag(pk) \\
    &=    \int_P \frac{\ud\mu_P(p)}{C^{w}}  \,\int_K \ud_{\texttt{haar}}(k)\,\int_P \ud\mu_P(p^{\prime})\,w(a(p^{\prime}))\,U(p)\, U(kp^{\prime}k^{-1})\,U^\dag(p) \\
    &=  \int_P \frac{\ud\mu_P(p)}{C^{w}}  \,\int_K \ud_{\texttt{haar}}(k)\,\int_P \ud\mu_P(k^{-1} p^{\prime}k)\,w(a(k^{-1} p^{\prime}k))\,U(p)\, U(p^{\prime})\,U^\dag(p) \\
    &= \int_P \frac{\ud\mu_P(p)}{C^{w}} \, \int_P \ud\mu_P(p^{\prime})\,w(a( p^{\prime}))\,U(p)\, U(p^{\prime})\,U^\dag(p) \\
    & = \int_P \frac{\ud\mu_P(p)}{C^{w}} \,{\sf M}^w(p)\,.
\end{align*}
The integral quantization of functions (or distributions) results from \eqref{residMw}
\begin{equation}
\label{ }
f(p) \mapsto A_f =\int_P \frac{\ud\mu_P(p)}{C^{w}} \, f(p) \,\sfM^{w}(p)\,.
\end{equation}

\section{SU$(1,1)$ quantizer operator from weight on the unit disk}
\label{SU11cIQ}
\subsection{Construction from weight}
The above material is now applied in the SU$(1,1)$ case.
Let us pick an $\eta > 1/2$ and choose a  function $[0,1] \ni u \equiv \vert z\vert^2 \mapsto w(u)$ such that the operator 
  \begin{equation}
\label{SU11Mw}
\MW= 2\int_{\D}\frac{\ud^2 z}{(1-\vert z\vert^2)^2}\, w(\vert z\vert^2) \, U^{\eta}(p(z))\,\sfP\, ,
\end{equation}
is bounded and  traceclass with unit trace, 
\begin{equation}
\label{unitrace}
\mathrm{tr}\,\MW = 1\,. 
\end{equation}
While comparing \eqref{SU11Mw} with the general construction \eqref{defMw}, one should notice the extra presence of twice the parity operator. The latter has been introduced here by convenience, as its importance will appear soon. Actually we could as well introduce its counterpart in  \eqref{defMw}, which is related to the above-mentioned inversion for Cartan symmetric domains. 

Supposing that we can invert infinite sum and integral, this condition together with \eqref{traceUnP}  imply the following normalisation of the weight function $w$:
\begin{equation}
\label{normw}
\int_{\D}\frac{\ud^2 z}{(1-\vert z\vert^2)^2}\, w(\vert z\vert^2)= 1\, . 
\end{equation}
Thus, if $w$ is non negative, it can be viewed as a probability distribution on the unit disk equipped with its SU$(1,1)$ invariant measure $\dfrac{\ud^2 z}{(1-\vert z\vert^2)^2}$.
From $(U^{\eta}(p(z)))^{\dag}= U^{\eta}(p(-z))$ and the invariance of the measure and $w$ under the change $z\mapsto -z$ one infers that $\MW$ is symmetric, 
and so self-adjoint.  Due to the isotropy of the weight function, $\MW$ is diagonal in the basis $\{e_n\}$, with matrix elements deduced from \eqref{matdiagpz} after the change $\vert z\vert \mapsto v=1-2\vert z\vert^2$:
 \begin{equation}
\label{matelMW}
\MW_{n n^{\prime}}= \delta_{n n^{\prime}}\,(-1)^{n} \,2^{2-\eta}\,\pi \,\int_{-1}^{1}\ud v\, w\left(\frac{1-v}{2}\right)\,(1+v)^{\eta-2}\,P_n^{(0,2\eta - 1)}(v)\, .
\end{equation}

\subsection{Resolution of the identity}

The unitarily transported versions $$\MW(p(z))= U^{\eta}(p(z)\, \MW\, U^{\eta}(p(-z))$$ of this operator are also bounded self-adjoint and are expected to resolve the unity with respect to a measure on $\D$ proportional to  $\ud^2 z/(1-\vert z \vert^2)^2$
\begin{equation}
\label{condMWresun}
I= \frac{1}{C^w}\,\int_{\D}\frac{\ud^2 z}{(1-\vert z\vert^2)^2} \, \MW(p(z))\, .
\end{equation}
 
 One  then computes $C^w$ with the simplest $\rho_0= |e_0\rg\lg e_0|$,
   \begin{equation}
\label{cste}
\begin{split}
C^w &= \int_{\D}\frac{\ud^2 z}{(1-\vert z\vert^2)^2}\,
\lg e_0 | \MW(p(z))|e_0\rg= \sum_n \MW_{nn}\,\int_{\D}\frac{\ud^2 z}{(1-\vert z\vert^2)^2}\, U^{\eta}_{0n}(p(z))\,\overline{U^{\eta}_{0n}(p(z))} \\
&= \sum_n \MW_{nn}\, \frac{\Gamma (2\eta + n)}{n!\,\Gamma(2\eta)}\,\int_{\D}\ud^2 z\,(1-\vert z\vert^2)^{2\eta -2}\, \vert z\vert^{2n} \\
&= \frac{\pi }{2\eta -1}\, ,
 \end{split}
 \end{equation}
 from the integral representation of the beta function and the unit trace of $\MW$. 
 Therefore, the resolution of the identity holds with the measure
\begin{equation}
\label{resunMW}
I= \frac{2\eta -1}{\pi}\, \int_{\D}\frac{\ud^2 z}{(1-\vert z\vert^2)^2} \, \MW(p(z))\, .
\end{equation}

\subsection{Particular weight functions}
 Let us consider  the particular family of positive weight functions
\begin{equation}
\label{partws}
w_s(u):= \frac{s-1}{\pi} \, (1-u)^s \, , \quad s>1\, , 
\end{equation}
which satisfy  \eqref{normw}.  
Using \eqref{JacobiInt}, we have for the matrix elements of $\sfM^{w_s;\eta}$
\begin{align}
\label{matelMW+sA}
\sfM^{w_{s};\eta}_{n n^{\prime}} &= (-1)^n\, \delta_{n n^{\prime}}\,  2(s-1)\,\frac{\Gamma(\eta +s-1)\Gamma(s-\eta)}{\Gamma(\eta +s +n)\Gamma(s-\eta -n)} \\
\label{matelMW+sB} &= \delta_{n n^{\prime}}\,  2(s-1)\, \frac{\Gamma(\eta +s-1)\Gamma(\eta -s + n+1)}{\Gamma(\eta +s +n)\Gamma(\eta -s +1)}\quad \mbox{for}\  s\neq \eta +1\, , \\
\label{matelMW+eta+1} & =\delta_{n n^{\prime}}\,\delta_{n 0}\, \quad \mbox{for}\  s= \eta +1\,.
\end{align}
From these expressions we see that the operator $\sfM^{w_{s};\eta}$ is a density operator if $1<s\leq \eta +1$. Positiveness is lost for $s > \eta +1$.  It is a finite rank $=p$ operator for all $s=\eta + p$, $p=1,2,\dotsc$.
The limit case \eqref{matelMW+eta+1}
 corresponds to Perelomov SU$(1,1)$ coherent states \cite{perel86} (with $\bar z$ instead of $z$):
\begin{equation}
\label{SU11CS}
\sfM^{w_{\eta + 1};\eta}=|e_0\rg\lg e_0| \, ,\quad  \sfM^{w_{\eta + 1};\eta}(p(\bar z)) = |z;\eta\rg
\lg z;\eta|\,,
\end{equation}
with 
\begin{equation}
\label{su11csper}
|z;\eta\rg  = (1 - \vert z \vert ^2)^{ \eta }\,  \sum_{n=0}^{\infty} \sqrt{\frac{(2 \eta)_n}{n!}} z^n |e_n \rg= U^{\eta} (p(\bar z))\, |e_0\rg\, .
\end{equation}
See  also the chapter 8 in \cite{gazeaubook09} for more details and specific properties. 

It is interesting to determine the weight function $w_{e_m;\eta}$ yielding the projector $|e_m\rg\lg e_m|$ through \eqref{SU11Mw}. By using the orthogonality relations satisfied by the Jacobi polynomials, we find 
\begin{equation}
\label{ememw}
\begin{split}
|e_m\rg\lg e_m|&= 2\int_{\D}\frac{\ud^2 z}{(1-\vert z\vert^2)^2}\, w_{e_m;\eta}(\vert z\vert^2) \, U^{\eta}(p(z))\, \sfP\, , \\ w_{e_m;\eta}(\vert z\vert^2)&= (-1)^m\,\frac{\eta +m}{\pi}\, (1-\vert z\vert^2)^{\eta+1}\,P_m^{(0,2\eta-1)}(1-2\vert z\vert^2)\,. 
\end{split}
\end{equation}
Clearly, at the exception of $m=0$, the weights $w_{e_m;\eta}$ are not non-negative. 

It is naturally possible to extend the range of values of $s$ below the limit $1$ at the price to violate integrability and positivity of the weight function.  As a matter of fact, there exists a remarkable value of $s$, namely $s=1/2$, for which the suitably renormalized weight function
\begin{equation}
\label{ws12}
w(u) = \frac{2\eta -1}{4\pi}\,(1-u)^{1/2}
\end{equation}
yields the identity operator,
\begin{equation}
\label{w12Id}
I= \sfM^{w_{1/2};\eta}= \frac{2\eta -1}{2\pi}\int_{\D}\frac{\ud^2 z}{(1-\vert z\vert^2)^{3/2}}\, U^{\eta}(p(z))\,\sfP\, ,
\end{equation}

From  the above we obtain  the integral representation of unit trace twice the parity operator:
\begin{equation}
\label{paritys12}
2\sfP 
 = \frac{2\eta -1}{\pi}\int_{\D}\frac{\ud^2 z}{(1-\vert z\vert^2)^{3/2}}\, U^{\eta}(p(z))\,.
\end{equation}
 As an interesting consequence of \eqref{paritys12}  combined with the resolution of the identity \eqref{condMWresun},  and 
 \begin{equation}
\label{PUPP}
U^\eta(p(z)) \sfP U^\eta(p(-z)) = U^\eta\left(p(z))^2\right) \sfP = U^\eta\left(p\left(\frac{2z}{1+\vert z\vert^2}\right)\right) \sfP, 
\end{equation}
 we obtain 
 \begin{equation}
\label{resunP}
I=2\, \frac{2\eta-1}{\pi}\,\int_{\D}\frac{\ud^2 z}{(1-\vert z\vert^2)^2} \, U^\eta\left(p\left(\frac{2z}{1+\vert z\vert^2}\right)\right) \sfP\, .
\end{equation}
 There results the (non trivial) integral formula given in \eqref{hyper} for hypergeometric polynomials.

\subsubsection*{Remark}

An open question is to  establish an inverse formula allowing to rebuild the weight $w$ from some trace formula, as it exists for the Weyl-Heisenberg or the affine case (see for instance \cite{gazeau18,bergaz18}).  More precisely, given $w$, the problem is to determine an operator $\mathfrak{I}$ such that  the following reconstruction formula holds. 
\begin{equation}
\label{dualw}
w(\vert z \vert^2) = \mathrm{tr}\,\left(\MW\,U^{\eta}(p(-z)\mathfrak{I}\right)\,. 
\end{equation}
\section{Weighted  SU$(1,1)$ integral quantizations for the unit disk}
\label{permissues}

We now start from the framework of the previous section and establish general formulas for the quantization issued from
a weight function $w(u)$  yielding the operator $\MW$ in \eqref{SU11Mw}:
\begin{equation}
\label{su11CIQ}
f\mapsto \sfAw_f= \frac{2\eta -1}{\pi}\,\int_{\D}\frac{\ud^2 z}{(1-\vert z\vert^2)^2} \,f(z)\, \MW(p(z))\, .
\end{equation}
Since the  operator $\sfAw_f$ acts on the Fock-Bargmann Hilbert space, the most straightforward way to characterize it is to compute its matrix elements with respect to the orthonormal basis \eqref{orthbasdisk2}. We know from \eqref{matelMW} that the operator $\MW$ is diagonal.
Hence, the general form of those matrix elements reads as
\begin{equation}
\label{su11CIQnnp}
\begin{split}
\left(\sfAw_f\right)_{n n^{\prime}}&= \frac{2\eta -1}{\pi}\int_{\D}\frac{\ud^2 z}{(1-\vert z\vert^2)^2} \,f(z)\, \MW(p(z))_{nn^{\prime}}\\
& = \frac{2\eta-1}{\pi}\sum_k\MW_{kk}\,\int_{\D}\frac{\ud^2 z}{(1-\vert z\vert^2)^2} \,f(z)\,
U^{\eta}_{nk}(p(z))\,\overline{U^{\eta}_{n^{\prime}k}(p(z))} \\
&:= \frac{2\eta-1}{\pi}\sum_k\MW_{kk}\,\mathcal{I}^{\eta}_{k,n,n^{\prime}}(f)\,. 
\end{split}
\end{equation}
The integral in the lhs has the explicit form derived from \eqref{Upz}
\begin{equation}
\label{Intknf}
\begin{split}
\mathcal{I}^{\eta}_{k,n,n^{\prime}}(f) &= \int_{\D}\frac{\ud^2 z}{(1-\vert z\vert^2)^2} \,f(z)\, 
U^{\eta}_{nk}(p(z))\,\overline{U^{\eta}_{n^{\prime}k}(p(z))}\\
&= \left( \frac{n_<!\, \Gamma(2 \eta + n_>) }{n_>!\, \Gamma(2 \eta + n_<)} \right)^{1/2} \,\left( \frac{n^{\prime}_<!\, \Gamma(2 \eta + n^{\prime}_>) }{n^{\prime}_>!\, \Gamma(2 \eta + n^{\prime}_<)} \right)^{1/2}\,(\mathrm{sgn}(n-k))^{n-k}\, (\mathrm{sgn}(n^{\prime}-k))^{n^{\prime}-k}\,  \times \\
&\times \int_{\D}\ud^2 z \,f(z) \left(1-\vert z \vert^2\right)^{2\eta -2} \, \vert z \vert^{n_> -n_< + n^{\prime}_> -n^{\prime}_<}\, e^{\ii(n^{\prime}-n)\phi}\times \\
& \times  P^{(n_> - n_<\, ,\, 2 \eta -1)}_{n_<}\left( 1-2\vert z\vert^2 \right)\,P^{(n^{\prime}_> - n^{\prime}_<\, ,\, 2 \eta -1)}_{n^{\prime}_<}\left( 1-2\vert z\vert^2 \right)\, , 
\end{split}
\end{equation}
with 
\begin{equation*}
n_{\substack{
>\\
<}}  = \left\lbrace \begin{array}{c}
    \max      \\
       \min
\end{array}\right.\, (n,k) \, ,  \quad n^{\prime}_{\substack{
>\\
<}}  = \left\lbrace \begin{array}{c}
    \max      \\
       \min
\end{array}\right.\, (n^{\prime},k)\,. 
\end{equation*}
In the isotropic case $f(z)= l(\vert z\vert^2)$ this integral simplifies to 
\begin{equation}
\label{Intknfis}
\begin{split}
\mathcal{I}^{\eta}_{k,n,n^{\prime}}(f)
&= \pi \delta_{nn^{\prime}}\,2^{1-2\eta + n_<-n_>} \,\frac{n_<!\, \Gamma(2 \eta + n_>) }{n_>!\, \Gamma(2 \eta + n_<)} \times\\
&\times \int_{-1}^{+1}\ud v \,l\left(\frac{1-v}{2}\right)\, (1-v)^{n_> -n_<} \, (1 +v)^{2\eta -2} 
\,  \left(P^{(n_> - n_<\, ,\, 2 \eta -1)}_{n_<}( v )\right)^2\, , 
\end{split}
\end{equation}
where we have used the variable $v= 1-2\vert z \vert^2$. Actually, it is sufficient to consider the case $n\leq k$, for which 
\begin{equation}
\label{Intknfisk}
\begin{split}
\mathcal{I}^{\eta}_{k,n,n^{\prime}}(f)
&= \pi \delta_{nn^{\prime}}\,2^{1-2\eta + n-k} \,\frac{n!\, \Gamma(2 \eta + k) }{k!\, \Gamma(2 \eta + n)} \times\\
&\times \int_{-1}^{+1}\ud v \,l\left(\frac{1-v}{2}\right) \, (1-v)^{k -n} \, (1 +v)^{2\eta -2}
\,  \left(P^{(k - n\, ,\, 2 \eta -1)}_{n}( v )\right)^2\, , 
\end{split}
\end{equation}
the case $k < n$ keeping  the same form thanks to the formula for Jacobi polynomials \cite{magnus66}, 
\begin{equation}
\label{jacobsym}
P_n^{( - a\, ,\, \beta)}( x )= \frac{\Gamma(n +\beta +1) \,(n-a)!}{\Gamma(n +\beta +1-a) \, n!}\, \left(\frac{x-1}{2}\right)\, P_{n-a}^{( a\, ,\, \beta)}( x )\quad \text{for} \quad a \in \N\,. 
\end{equation}
Of course, for $f=1=l$, we should recover the identity in \ref{su11CIQnnp}, which implies the following value for the integral $\mathcal{I}^{\eta}_{k,n,n^{\prime}}(1)$ (which can be found in \cite{gradryz07}),
\begin{equation}
\label{Ikn1}
\begin{split}
\mathcal{I}^{\eta}_{k,n,n^{\prime}}(1)
&= \pi \delta_{nn^{\prime}}\,2^{1-2\eta + n-k} \,\frac{n!\, \Gamma(2 \eta + k) }{k!\, \Gamma(2 \eta + n)} \times\\
&\times \int_{-1}^{+1}\ud v  \, (1-v)^{k -n} \,(1 +v)^{2\eta -2}
\,  \left(P^{(k - n\, ,\, 2 \eta -1)}_{n}( v )\right)^2\\
&= \frac{\pi}{2\eta - 1}\, . 
\end{split}
\end{equation}
Another useful particular case is $f(\vert z\vert^2) = (1-\vert z\vert^2)^{-1}= 2/(1+v)$. Then, 
\begin{equation}
\label{Ikn2}
\begin{split}
\mathcal{I}^{\eta}_{k,n,n^{\prime}}\left(\frac{1}{1-\vert z\vert^2}\right)
&= \pi \delta_{nn^{\prime}}\,2^{2-2\eta + n-k} \,\frac{n!\, \Gamma(2 \eta + k) }{k!\, \Gamma(2 \eta + n)} \times\\
&\times \int_{-1}^{+1}\ud v \, (1-v)^{k -n} \, (1 +v)^{2\eta -3} 
\,  \left(P^{(k - n\, ,\, 2 \eta -1)}_{n}( v )\right)^2\\
&= \delta_{nn^{\prime}}\,\frac{\pi}{2\eta - 1}\,\frac{1}{2\eta \,(\eta-1)}[(k+\eta)(\eta + n) + \eta(\eta -1)]\, . 
\end{split}
\end{equation}
This formula is derived from the equation \eqref{jac22}.

By construction, the quantization map \eqref{su11CIQ} is covariant with respect to the unitary
action $U^{\eta}$ of SU$(1,1)$:
\begin{equation}
\label{covsu11} U^{\eta}(g_0) \, \sfAw_f \,{U^{\eta}}^{\dag}(g_0) =
\sfAw_{\mathfrak{U}(g_0)f}\, ,
\end{equation}
where we recall that $(\mathfrak{U}(g)f)(z)= f\left(g^{-1}\cdot z\right)$. 
Moreover, due to the self-adjointness of $\MW(p(z))$ we have the relation
\begin{equation}
\label{symAf}
(\sfAw_f)^{\dag} = \sfAw_{\bar{f}}\, \,. 
\end{equation}
The following important statement   results from  the covariance  \eqref{covsu11} and self-adjointness property \eqref{symAf}.
\beprop
\label{qk0pm}
Suppose that $w$ and $\eta > 1/2$ are such that the  series 
\begin{equation}
\label{serkm}
\mathcal{S}^{w;\eta}:=\sum_{k=0}^{\infty} k\, \sfM^{w;\eta}_{k k} 
\end{equation}
converges in a certain sense. 
Then the quantization \eqref{su11CIQ} maps the basic observables $k_a$, $a=0,1,2$ (resp. $a=0,\pm$), defined in \eqref{3classobs}, to the self-adjoint generators \eqref{uirepsu110}, \eqref{uirepsu111}, \eqref{uirepsu112}, up to a constant factor $\gamma_{w;\eta}$
\begin{equation}
\label{quantgen}
\sfAw_{k_a} = \gamma_{w;\eta}\,K_{a}\, , \quad a=0,1,2\, , \quad \text{resp.} \ a=0,\pm\, , 
\end{equation}
with
\begin{equation}
\label{gamweta}
\gamma_{w;\eta}= \frac{1}{\eta - 1}\left[ 1 + \frac{\mathcal{S}^{w;\eta}}{\eta}\right]\,. 
\end{equation}
\enprop
\bprf
We have from \eqref{su11CIQ}
\begin{equation}
\label{Akaka}
A_{k_a} = \frac{2\eta -1}{\pi}\,\int_{\D}\frac{\ud^2 z}{(1-\vert z\vert^2)^2} \,k_a(z)\, \MW(p(z))\, . 
\end{equation}
From the covariance \eqref{covsu11} and from \eqref{Umatekab}
\begin{equation}
\label{covsu11ka} U^{\eta}(g) \, \sfAw_{k_a} \,{U^{\eta}}^{\dag}(g) = \sfAw_{\mathfrak{U}(g)k_a}= \sum_b [\mathfrak{U}(g)]_{ba}\, \sfAw_{k_b} 
\quad \forall \, g\in \text{SU}(1,1)\, .
\end{equation}
This means that the operators $\sfAw_{k_a}$ transform under the action of $U^{\eta}(g)$ exactly like the operators $K_a$. Thus, there exists a constant $\gamma_{w;\eta}$ depending on $\eta$ and $w$ such that
\begin{equation}
\label{AkaKa}
\sfAw_{k_a} = \gamma_{w;\eta} K_a\, . 
\end{equation}
The constant is calculated by picking $a=0$ and  considering the lowest  matrix elements via \eqref{Ikn2},
\begin{equation}
\label{constgam}
\begin{split}
\gamma_{w;\eta}\left(K_0\right)_{00}&=\gamma_{w;\eta}\, \eta= \left(\sfAw_{k_0}\right)_{00} = \frac{2\eta-1}{\pi}\sum_k\MW_{kk}\,\mathcal{I}^{\eta}_{k,0,0}(k_0)\\
&= 2\frac{2\eta-1}{\pi}\sum_k\MW_{kk}\,\mathcal{I}^{\eta}_{k,0,0}\left(\frac{1}{(1-\vert z\vert^2}\right) - 1 \\
&= \frac{1}{\eta -1}\left[ \eta + \mathcal{S}^{w;\eta}\right]\, ,
\end{split}
\end{equation}
we finally get \eqref{gamweta}. 

Once it is proved for $\sfAw_{k_0}$, it is also proved for $\sfAw_{k_\pm}$ by using \eqref{covsu11ka} with specific elements $g$'s mapping $K_0$ to $K_\pm$. 
\eprf
For instance, if we choose $w= w_s$ given by \eqref{partws} with $s= \eta +1$ (Perelomov CS case), the formula \eqref{matelMW+eta+1} gives $\mathcal{S}^{w_{\eta + 1};\eta}=0$ and so 
\begin{equation}
\label{gamCSPer}
\gamma_{w_{\eta + 1};\eta}= \frac{1}{\eta - 1}\, .
\end{equation}
We note that with the particular value $\eta = 2$, the quantization of basic observables is exact.  
In the case of coherent states built from $|e_m\rg$  and giving rise to \eqref{ememw}, the constant $\gamma_{w;\eta}$ is given by
\begin{equation}
\label{gammaem}
\gamma_{w_{e_m};\eta}= \frac{\eta + m}{\eta(\eta-1)}\,. 
\end{equation} 
Another interesting case related to $w= w_s$ concerns the already mentioned  limit value $s= 1/2$ for which $\mathcal{S}^{w_{1/2};\eta}=2\sum_{k=0}^{\infty} k\,(-1)^{k} = -1/2$ (in Abel sense). Then
\begin{equation}
\label{gamweyl}
\gamma_{w_{1/2};\eta}= \frac{2\eta -1}{2\eta(\eta-1)}\,,
\end{equation}
and there is no real value of $\eta$ for which the quantization of the basic observables is exact.

Finally, the interesting functions to be quantized have the general form $f(z)=h(\vert z\vert^2)z^a$, $a\in \N$, where $h$ is real-valued. Note that the  quantization of the conjugate is straightforward, due to the relation \eqref{symAf}. 
However, in view of the technicality of the calculations, we will not pursue in this way. 


\section{Quantum phase space portraits}
\label{semclass}

Let us consider a weight function $w(\vert z\vert^2)$ yielding the symmetric unit trace operator $M^{w;\eta}$ through Eq. \eqref{SU11Mw}.  The semi-classical or lower symbol of an operator $A$ in $\mathcal{H}$ is the function
\begin{equation}
\label{lowsymbA}
\check{A}(z):=
\mathrm{tr}\,\left(A\,U^{\eta}(p(z))\,\MW\,\left(U^{\eta}(p(z)\right)^{\dag}\right)
=\mathrm{tr}\,\left(A\,\MW(p(z))\right)\,.
\end{equation}
Let us now   consider a function $f(z)$ and its quantum version ${\mathsf{A}}^{w_1;\eta}_f$ built from a first weight function $w_1(\vert z\vert^2)$, used for the ``analysis". Its lower symbol associated with a second weight function $w_2(\vert z\vert^2)$, used for the ``reconstruction'' (terms borrowed from signal analysis terminology) reads as   the map 
\begin{equation}
\label{ffch1}
\begin{split}
f(z) \mapsto &\check{f}(z)\equiv \check{\mathsf{A}}^{w_{12};\eta}_f(z)\\
&=\frac{2\eta-1}{\pi}\, \int_{\D}
\frac{\ud^2z^{\prime}}{(1-\vert z^{\prime}\vert ^2)^{2}}\, f(z^{\prime})\,
\mathrm{tr}\,\left(\MWF(p(-z)p(z^{\prime}))\,\MWS \right)\,.
\end{split}
\end{equation}
Now, we have the SU$(1,1)$ composition formula, 
\begin{equation}
\label{pzpzp}
p(-z)p(z^{\prime})= p(t)\,h(\theta)\, , \ \text{with}\ t= p(-z)\cdot z^{\prime}\, , 
\end{equation}
and $h(\theta)\in H$. Since $U^{\eta}(h(\theta))$ is diagonal, it commutes with $\MW$, and we obtain after the change of variable $z^{\prime}\mapsto t$, 
\begin{equation}
\label{ffch2}
\check{f}(z)=\frac{2\eta-1}{\pi}\, \int_{\D}
\frac{\ud^2t}{(1-\vert t\vert ^2)^{2}}\,\, f\left(p(z)\cdot t\right)\,\mathrm{tr}\,\left(\MWF(p(t))\,\MWS \right)\,.
\end{equation}
Clearly, since for $f=1$ the rhs is equal to $1$, the map 
\begin{equation}
\label{probdist}
t\mapsto \mathrm{tr}\,\left(\MWF(p(t))\,\MWS\right) = \mathrm{tr}\,\left(U^{\eta}(p(t))\MWF\,\left[\MWS U^{\eta}(p(t))\right]^{\dag}\right) 
\end{equation}  
is a probability distribution on the unit disk $\D$ with respect to the measure $\dfrac{2\eta-1}{\pi}\dfrac{\ud^2t}{(1-\vert t\vert ^2)^{2}}$ if $\MWF$ and $\MWS$ are nonnegative, i.e., are density operator. 

Let us just prove that the lower symbols of the three generators $K_0$, $K_{\pm}$, are proportional to their classical counterpart:
\begin{equation}
\label{lowsymbgen}
\check{\mathsf{A}}^{w_{12};\eta}_{k_i}(z)=\varkappa_{w_{12};\eta}k_i(z)\,, \quad i=0, 1,2\,.  
\end{equation}
The proof is similar to the one of Proposition \ref{qk0pm}. Let us apply the regular representation of SU$(1,1)$ on both sides of Eq. \eqref{ffch2}. 
\begin{equation*}
\label{}
\check{f}\left(g^{-1}\cdot z\right)= \frac{2\eta-1}{\pi}\, \int_{\D}
\frac{\ud^2t}{(1-\vert t\vert ^2)^{2}}\,\, f\left(p(g^{-1}\cdot z)\cdot t\right)\,\mathrm{tr}\,\left(\MWF(p(t))\,\MWS \right)\,.
\end{equation*}
We now apply \eqref{leftactSU11}:
\begin{equation*}
\label{ }
p(g^{-1}\cdot z)\cdot t= \left(g^{-1}\,p(z)\,h\right)\cdot t = g^{-1}\cdot(p(z)\cdot(h\cdot t))\, ,\quad \mbox{with}\quad h\in \mathrm{U}(1)\, . 
\end{equation*}
Hence, after changing $h\cdot t \mapsto t$ and using the invariance or the measure and of the trace, we get
\begin{equation*}
\label{}
\check{f}\left(g^{-1}\cdot z\right)= \frac{2\eta-1}{\pi}\, \int_{\D}
\frac{\ud^2t}{(1-\vert t\vert ^2)^{2}}\,\, f\left(g^{-1}\cdot (p(z)\cdot t)\right)\,\mathrm{tr}\,\left(\MWF(p(t))\,\MWS \right)\,.
\end{equation*}
Hence, by particularizing to $f=k_a$, $a=0,\pm$, we check by linearity that their corresponding  $\hat k_a$ transform exactly in the same way as in  \eqref{k0g},\eqref{k+g},\eqref{k-g},  under the regular representation of SU$(1,1)$. This proves the proportionality relation \eqref{lowsymbgen}. The constant is computed by using a similar trick to \eqref{constgam}. Of course, these formulae are valid only if $\eta$ and  the weights $w_1$, $w_2$ are such that the integral \eqref{ffch2} converges for each one of the considered cases. As an elementary example let us choose $w_1(u)=w_2(u)=(\eta/\pi)(1-u)^{\eta +1}$,  which corresponds to the Perelomov case \eqref{SU11CS}. 
Using the transformations \eqref{k0p},\eqref{k+p}, and \eqref{k-p},  we find for \eqref{ffch2}
\begin{align*}
\check{f}_a(z)&=\frac{2\eta-1}{\pi}\, \int_{\D}
\frac{\ud^2t}{(1-\vert t\vert ^2)^{2}}\, k_a(\left(p(z)\cdot t\right)\,\vert \lg t;\eta|0;\eta\rg\vert^2\\ & =\frac{2\eta_1-1}{\pi}\, \int_{\D}
\ud^2t\,(1-\vert t\vert ^2)^{2\eta  -2}\, k_a(\left(p(z)\cdot t\right)\\
&= k_a(z)\,\left[\frac{2\eta-1}{\pi}\, \int_{\D}
\ud^2t\,(1-\vert t\vert ^2)^{2\eta  -2}\, k_0( t)\right]\\
&= \frac{2\eta}{\eta -1}\,k_a(z)\,. 
\end{align*}
Note that this proportionality  coefficient cannot be put equal to $1$ for the allowed range $\eta >1$.
An interesting problem is to choose $\eta = 2$, $w_1(u)=(\eta/\pi)(1-u)^{\eta +1}= 2\pi(1-u)^{3}$ which corresponds to a Perelomov case, and which yields the exact quantization for the 3 basic observables, and to build the reconstruction operator $\MWS$ which yields exactly $\check{f}_a(z)= k_a(z)$. 

\section{Conclusion}
\label{conclu}
Given an irreducible unitary representation $U^{\eta}$, $\eta >1$ in the discrete series of SU$(1,1)$, we have presented a family of covariant integral quantizations of functions or distributions on the open unit disk.  A physical interpretation is to consider SU$(1,1)$ as the kinematical group of the $1+1$ AdS space-time and the disk as the phase space for the motion of a ``massive"  Wigner elementary system in AdS.  Each quantization is determined by an  isotropic weight function on the disk, or equivalently by a unit trace class not necessarily positive operator viewed as a ``$U^{\eta}$-Fourier transform'' of this weight. Perelomov coherent states quantizations are particular cases.   Reversal of these quantizations under the form of semi-classical portraits of quantized versions of a classical object  have been defined as local averaging of the latter, involving a second weight function. In this regard, a non-trivial  question to be considered is to determine a pair $(w_1,w_2)$ of weight functions for which the reversal is exact in the Wigner-Weyl sense, i.e., the following holds
\begin{equation}
\label{ffch11}
f(z) \mapsto \check{f}(z)\equiv \check{\mathsf{A}}^{w_{12};\eta}_f(z)= f(z)\, , 
\end{equation}
with the notations of \eqref{ffch1}. To a certain extent, this problem could be viewed as a generalisation to SU$(1,1)$ of
similar approaches concerning the Weyl-Heisenberg group, the affine group, and more general groups and their related Weyl operators and Wigner functions defined  in a wide sense, see for instance \cite{cohen66,cohen13,agawo70,gazeau18,bergaz18,alatchuwo00}, and reference therein. 
 
Finally, appealing generalisations of the presented material are for the higher-dimensional  Anti de Sitter groups, since some of these groups might have physically relevant discrete series.  


\appendix

\section{Useful integrals with Jacobi polynomials and others}
\label{jacobiint}
\subsubsection*{Orthogonality}
\begin{equation}
\label{jacortho}
\begin{split}
&\int_{-1}^{+1}\ud v \, (1-v)^{\alpha} \, (1 +v)^{\beta} \,  P^{(\alpha\, ,\, \beta)}_{m}( v )\, P^{(\alpha\, ,\, \beta)}_{n}( v )\\&=\delta_{mn}\, \frac{2^{\alpha + \beta +1}}{\alpha + \beta + 2n+1}\,\frac{\Gamma(\alpha + n +1)\,\Gamma(\beta + n +1)}{n!\,\Gamma(\alpha +\beta + n +1)}\, , 
\end{split}
\end{equation}
for $\alpha > -1$ and $\beta >- 1$.

\subsubsection*{Others}
From Gradshteyn-Ryzhik 7.391 in \cite{gradryz07}
\begin{equation}
\label{jac12}
\begin{split}
&\int_{-1}^{+1}\ud v \, (1-v)^{\alpha} \, (1 +v)^{\beta-1} \,  \left(P^{(\alpha\, ,\, \beta)}_{n}( v )\right)^2\\
&=\frac{2^{\alpha + \beta}}{ \beta}\,\frac{\Gamma(\alpha + n +1)\,\Gamma(\beta + n +1)}{n!\,\Gamma(\alpha +\beta + n +1)}\,,
\end{split}
\end{equation}
for $\alpha > -1$ and $\beta > 0$. 

A new one. 
\begin{equation}
\label{jac22}
\begin{split}
&\int_{-1}^{+1}\ud v \, (1-v)^{\alpha} \, (1 +v)^{\beta-2} \,  \left(P^{(\alpha\, ,\, \beta)}_{n}( v )\right)^2\\
&=\frac{2^{\alpha+\beta-1}}{\beta(\beta+1)(\beta-1)}\,\frac{(\Gamma(\alpha+n+1)!\,\Gamma(\beta + n+1)}{n!\, \Gamma(\alpha + \beta +  n+1) }\, [(\beta +1)(\alpha + \beta) + 2(\alpha + \beta +n+1)n]\, , 
\end{split}
\end{equation}
for $\alpha > -1$ and $\beta > 1$. 

From Gradshteyn-Ryzhik 7.391 in \cite{gradryz07}
\begin{align}
\label{JacobiInt}
\nonumber \int_{-1}^1\ud x (1-x)^{\rho}\, (1+x)^{\sigma}\, P_n^{(\mu,\nu)}(x)&= 2^{\rho + \sigma+1}\,\frac{\Gamma(\rho + 1)\,\Gamma(\sigma + 1)\,\Gamma(n + 1+ \mu)}{n!\, \Gamma(\rho +\sigma+ 2)\,\Gamma(\mu + 1)}\times\\
&\times {}_3F_2(-n,\mu + \nu+n+1, \rho+1; \mu+1,\rho+\sigma + 2;1)\, ,
\end{align}
with $\mathrm{Re}\,\rho > -1\, ,\mathrm{Re}\,\sigma > -1$

\subsubsection*{A new integral for hypergeometric polynomials}
\begin{equation}
\label{hyper}
2\,(2\eta-1)\int_0^1\ud u\,(1-u)^{2\eta-2}\,(1+u)^{-2\eta}\, {}_2F_1\left(-n,n+2\eta;1; \frac{4u}{(1+u)^2}\right)= (-1)^n\,. 
\end{equation}

\subsubsection*{Two integral forms for beta function }
\begin{equation}
\label{betaforms}
\begin{split}
\beta(x,y)&= \frac{\Gamma(x)\,\Gamma(y)}{\Gamma(x+y)}=\int_0^1\ud t \, t^{x-1}\,(1-t)^{y-1}\\
&= 2^{1-x-y}\int_{-1}^1\ud t \, (1-t)^{x-1}\,(1+t)^{y-1}\,. 
\end{split}
\end{equation}

\subsection*{Acknowledgments} This research is supported in part by the Ministerio de Econom\'ia y Competitividad of Spain  under grant  MTM2014-57129-C2-1-P and the Junta de Castilla y Le\'on (grant VA137G18). J.P.G. is also indebted to the University of Valladolid for its hospitality.

\end{document}